\documentclass[12pt]{article}
\usepackage{amscd,amsmath,amssymb,amstext,amsthm,exscale,latexsym}
\usepackage{graphicx}
\usepackage{upgreek,txfonts}
\newcommand {\al}   {\alpha}       \newcommand {\bt}  {\beta}
\newcommand {\g }   {\gamma}       
\newcommand {\dl}   {\delta}

\newcommand {\vf }  {\varphi}      
         
\newcommand {\Lm}   {\Lambda}      \newcommand {\Om}  {\Omega}
       
\newcommand {\pl}   {\partial}

\renewcommand {\ln}{{\sf\,ln}}         

\newcommand {\vol}  {\sqrt{{\scriptstyle |}g{\scriptstyle |}}}
            
   \newcommand {\MR}  {{\mathbb R}}

\begin{document}
\title     {Black hole formation by a scalar field}
\author    {D. E. Afanasev
            \thanks{E-mail: daniel\_afanasev@yahoo.com}\\
            \sl Moscow Aviation Institute (National Research University),
            \\
            \sl Volokolamskoe shosse, 4, Moscow, 125993, Russia
            \\ M. O. Katanaev
            \thanks{E-mail: katanaev@mi-ras.ru}\\
            \sl Steklov mathematical institute,
            \sl ul.~Gubkina, 8, Moscow, 119991, Russia}
\date      {}
\maketitle
\begin{abstract}
The Liouville solution in General Relativity with a scalar field is discussed.
This solution is invariant with respect to global Lorentz transformations, and
dependence on time cannot be removed. If the scalar field potential is
exponential and unbounded from below, the Liouville solution describes the
formation of spherically symmetric black hole. The event horizon is a sphere,
which appears with infinitesimal radius at a finite moment of time and
afterwards expands with the velocity of light to infinity. A distant observer
can measure the geometric defect at the point where the horizon appears. It is
similar to the defect produced by the monopole or spherical dislocation of
space-time. Comparison with the Schwarzschild solution yields the mass function
which is proportional to the time squared.
\end{abstract}
{\em Introduction.}
The problem of a gravitational collapse and subsequent black hole formation is one
of the most important and intriguing in theoretical physics. Extensive research
started from the seminal paper by J.~R. Oppenheimer and H.~Snyder \cite{OppSny39}, where
the authors considered a spherically symmetric collapse of fluid-like matter of
a star and supposed that the outside metric is the Schwarzschild one. These
assumptions are physically reasonable, but there is the problem of smooth
gluing of exterior and interior regions of the space-time. Mathematically, it is
not a solution of Einstein's equations on the whole space-time because they are
not satisfied on the gluing surface.

Gravitational collapse of matter fields filling the whole space-time is also of great interest. In particular, models with a scalar field provide a natural testing ground for understanding of the process. On this path, massless scalar field models have been extensively studied both analytically and numerically as a simplest framework for a gravitational collapse \cite{Christo91,Choptu93}. Subsequent investigations involve various scalar field potentials \cite{Heard02,Chakra17,Mosa23,CorGiam24}, including negative exponential form playing a central role in the ekpyrotic scenario where such potentials drive cosmic contraction \cite{Lehn08}. For a more detailed introduction to the scalar field collapse, we direct the reader to the recent review \cite{Giamb23} and the references therein.   

Despite the extensive literature, exact global solutions where interior region of the black hole smoothly matched to an exterior geometry remain scarce. In the present paper, we show that the recently found Liouville solution
\cite{AfaKat25A,AfaKat25B} describes the formation of a spherically symmetric
black hole by a scalar field. The solution is global and infinitely smooth in the whole space-time, that is the field equations are satisfied everywhere. The event
horizon is a two-dimensional sphere, which appears with infinitesimal radius at
a finite moment of time and afterwards expands to infinity with the velocity of
light. Later the spacelike singularity appears that is completely hidden by the horizon.

The interior of the black hole is a homogeneous and isotropic
region with negative spatial curvature like the Friedmann spacetime, the
horizon being the light cone corresponds to the zero of the scale factor.
It is known \cite{Lin18}, that, under some assumptions, a hypersurface of a vanishing scale factor
is geodesically incomplete and non-singular for Friedmann space-times with
negative spatial curvature as in our case. So, the
solution we found capture both the scalar field collapse inside the Friedmann
space-time and the smooth extension of geodesics beyond the zero of the scale
factor into the external region where isotropy and homogeneity are broken.

The space-time geometry outside the black hole is not Schwarzschild-like, but
can be compared to it at large distances. The respective mass function is
proportional to time squared divided by radius. Moreover, the analysis of the asymptotic behavior of the spacetime metric reveals the presence of a geometric defect similar to the global monopole \cite{BarVil89}.  

The Liouville solution is obtained for the specific exponential potential for a
scalar field and may serve at least
as a toy model, because the strategy is straightforward: the action $\to$ the
field equations $\to$ the Liouville ansatz for the metric $\to$ exact smooth
global solution without any approximation.


{\em The Liouville solution with the black hole.}
We consider four-dimensional space-time equipped with the Lorentz signature
metric $g$. Local coordinates are denoted by $x=(x^\al)$, where
$\al=0,1,2,3$. The action of the model is
\begin{equation}                                                  \label{ancsgt}
  S:=\int\!dx\vol\left[R-2\Lm+\frac12g^{\al\bt}\pl_\al\vf\pl_\bt\vf-V(\vf)
  \right],\qquad g:=\text{det} g_{\al\bt},
\end{equation}
where $\vf(x)$ is a scalar field, $\Lm$ is a cosmological constant, and $V$ is
a scalar field potential. The field equations were solved
\cite{AfaKat25A,AfaKat25B} using the conformally flat Liouville-type metric
\cite{Liouvi49}
\begin{equation*}
  g_{\al\bt}:=\Phi(x)\eta_{\al\bt},\qquad\eta_{\al\bt}:=\text{diag}(+1,-1,-1,-1),
\end{equation*}
where the conformal factor $\Phi$ is the sum of four functions on single
arguments
\begin{equation*}
  \Phi(x):=\phi_0(x^0)+\phi_1(x^1)+\phi_2(x^2)+\phi_3(x^3),
  \qquad\Phi>0.
\end{equation*}
It was proved that a general solution of the field equations exists only for the
exponential potential
\begin{equation}                                                  \label{qnshdf}
  V+2\Lm=\pm 12C^2\text{e}^{k\vf},\qquad C=\text{const}>0,
\end{equation}
where $k=\pm2/\sqrt{3}$ independently of the sign in front of the exponent. If
the potential is bounded from below \big(the plus sign in Eq.~(\ref{qnshdf})\big)
a general solution for the metric has the form
\begin{equation}                                                  \label{endhgf}
  g_{\al\bt}=\Phi(s)\eta_{\al\bt}=(c+s)\eta_{\al\bt},\qquad
  s:=\eta_{\al\bt}x^{\al} x^{\bt},\qquad c\in\MR.
\end{equation}
This metric is defined and infinitely smooth in the domain $s>-c$ and
non-trivially depends both on time $x^0$ and spatial coordinates. It is
explicitly invariant under global Lorentz rotations about the origin and,
in particular, is spherically symmetric. Analysis of geodesics \cite{AfaKat25B}
shows that this solution is a global one: any geodesic can be either continued
to infinite value of the canonical parameter in both directions or it ends up a
singular point where both curvature and scalar field diverge.

For $c>0$ a general solution describes the naked singularity and has
cosmological interpretation. That is the solution inside the light cone with the
vertex at the origin describes homogeneous and isotropic Universe with
accelerated expansion. Outside the light cone the solution becomes
nonhomogeneous and nonisotropic, and contains naked singularity which attracts
matter from inside the light cone thus providing accelerated expansion of the
Universe without need to introduce additional dark energy. Furthermore, the
cosmological scale factor vanishes on the whole light cone but there is no
singularity in curvature and scalar field. The Liouville solution shows
explicitly that the cosmological model with homogeneous and isotropic Universe
can be smoothly continued through the zero of the scale factor which is located
not at a point but on the whole light cone.

The aim of the present paper is to analyse the solution of the model with action
(\ref{ancsgt}) for the potential bounded from above \big(the minus sign in
Eq.~(\ref{qnshdf})\big). Though the unbounded from below potential has severe
problems in quantum field theory, we consider this case for two reasons: (i)
unbounded potentials play important role in classical mechanics (for example,
the Newton potential) and (ii) the Liouville solution for the potential bounded
from above and unbounded from below describes very important physical process.
Namely, there is a solution describing spherically symmetric black hole
formation by the scalar field.

We do not repeat the derivation of a general solution
\cite{AfaKat25A,AfaKat25B}. In the case under consideration, the space-time
metric for the Liouville solution differs from that in Eq.~(\ref{endhgf}) only
by the sign of $s$:
\begin{equation}                                                 \label{solmetr}
  g_{\al\bt}=\Phi(s)\eta_{\al\bt}=(c-s)\eta_{\al\bt}.
\end{equation}
The case of interest is $c>0$, which is considered. The scalar field has a
logarithmic form and depends on coordinates only through the conformal factor
\begin{equation*}
  \vf=-\frac2k\ln\big(C\Phi\big),
\end{equation*}
for both signs in Eq.~(\ref{qnshdf}). It is constant on hyper-surfaces
$s=\text{const}<c$ and diverges as $s\to -\infty$ and $s\to c$.

The curvature tensor for metric~(\ref{solmetr}) is
\begin{equation*}                                               \label{curvature}
\begin{split}
  R_{\al\bt\g}{}^\dl=&\frac1\Phi\left(-3+\frac c\Phi\right)
  \big(\eta_{\al\g}\dl_\bt^\dl-\eta_{\bt\g}\dl_\al^\dl)-
\\
  &-\frac3{\Phi^2}\big(x_\al x_\g\dl_\bt^\dl-x_\bt x_\g\dl_\al^\dl
  -x_\al x^\dl\eta_{\bt\g}+x_\bt x^\dl\eta_{\al\g}\big).
\end{split}
\end{equation*}
Its components tend to zero for $s\to -\infty$. However, since metric
(\ref{solmetr}) becomes degenerate in this limit, one cannot claim that the
space-time is asymptotically Minkowskian. Furthermore, we shall see that the
solution describes a geometric defect at the centre similar to the global
monopole \cite{BarVil89} or space-time spherical dislocation with an excess of
the solid angle within the geometric theory of defects \cite{KatVol92,Katana05}.
The scalar curvature
\begin{equation*}
  R=-\frac{6}{\Phi^2}\left(3+\frac c\Phi\right)
\end{equation*}
is singular for $\Phi=0\Leftrightarrow s=c$. The singularity is space-like and
forms two-sheeted hyperboloid for $c>0$.

\begin{figure}[hbt]
\hfill\includegraphics[width=.35\textwidth]{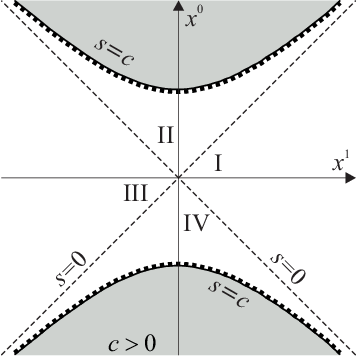}
\hfill {}
\centering\caption{The domain of the Liouville solution (unshaded region) for
the black hole formation on the $x^0$, $x^1$ slice of the space-time. The dotted
and zigzag lines denote, respectively, the light cone which is the event horizon
and the singularity.}
\label{Liouville-1}
\end{figure}

Solution (\ref{solmetr}) is global, infinitely smooth and defined on the one
sheeted hyperboloid as shown in Fig.~\ref{Liouville-1}.
Note, that the numbering of quadrants is purely formal because the picture
must be rotated in two extra space directions, and therefore quadrants I and III
represent a single connected Universe (external region). Since the space-time is
conformally flat, its causal
structure coincides with that of Minkowski space-time. The zigzag lines denote
black and white holes, where both curvature and scalar field become singular,
the singularity being geodesically incomplete. The future light cone $s=0$,
$x^0\ge0$ is the event horizon of the black hole: no signal emitted from
quadrant II (under horizon) can reach future null infinity. There are no time
infinities.

Figure \ref{Liouville-1} resembles that of the Schwarzschild solution, but there
is an important difference: as noted above, we have only one Universe, unlike
the Schwarzschild case, where quadrants I and III represent two different
causally disconnected regions of space-time. We also note, that inside quadrants II and IV metric~(\ref{solmetr}) can be brought to the Friedmann form and, therefore, they represent homogeneous and isotropic regions as in the case of the Liouville solution with the naked singularity mentioned above. 

Given the spherical symmetry, it is reasonable to introduce spherical
coordinates $(t,r,\theta,\vf)$ and express metric~(\ref{solmetr}) in the form
\begin{equation}\label{metrbb}
  ds^2=\left(c-t^2+r^2)\right)\left(dt^2-dr^2-r^2d\Omega\right),
\end{equation}
where $d\Omega$ is the metric of the unit two-dimensional sphere. Note, that the
limit $s\to -\infty$ implies $r\to+\infty$, that corresponds to spatial
infinity. In the limit $r\to\infty$, $t/r\to 0$, the respective asymptotic of
metric~(\ref{metrbb}) to leading order is
\begin{equation*}
  ds^2=r^2\big(dt^2-dr^2-r^2d\Omega\big),
\end{equation*}
It can be checked, that curvature of this metric is nonzero. It is the
consequence of the coordinate degeneracy of the metric at spatial infinity.

In spherical coordinates, the event horizon equation reads
\begin{equation*}
  t^2-r^2=0,\qquad t>0.
\end{equation*}
For fixed $t>0$, the event horizon is a spatial two-sphere of finite radius
$r=t$, which emerges with infinitesimal size at $t=0$ and expands with the
velocity of light during time evolution. The spacelike singularity forms at
$t=c$ and is completely hidden inside the horizon. We see that the Liouville
solution for the scalar field potential unbounded from below describes the
process of the black hole formation by the scalar field.

{\em Geometric defect.}
To analyze the asymptotic behavior of the Liouville solution at spatial
infinity we introduce new coordinates:
\begin{equation}                                                   \label{TtRr}
  R:=\frac12 r^2,\qquad T:=tr.
\end{equation}
Note, that equality $T/R=2t/r$ shows that the limits ($R\to\infty$, $T/R\to 0$)
and ($r\to\infty$, $t/r\to 0$) are equivalent.

In coordinates~(\ref{TtRr}), the event horizon equation reads:
\begin{equation} \label{evhor}
  T-2R=0,\qquad T>0.
\end{equation}
Metric~(\ref{metrbb}) takes the form:
\begin{equation}                                                  \label{metrTR}
  ds^2=\left(1+\frac c{2R}-\frac{T^2}{4R^2}\right)
  \left[\left(dT-\frac{T}{2R}dR\right)^2-dR^2-4R^2d\Omega\right],
\end{equation}
At spatial infinity $R\to\infty$ for $T/R\to 0$ it tends to
\begin{equation*}
  ds^2=dT^2-dR^2-4R^2d\Omega.
\end{equation*}
This limiting metric contains an extra factor 4 in front of the angular part
compared to the Lorentz metric, which cannot be eliminated by coordinate
transformations. The geometric difference is essential, because the area of the
two-sphere now equals $16\pi R^2$. This circumstance indicates the presence of
a geometric defect similar to the one which is present due to the monopole
\cite{BarVil89}. Unlike the standard monopole case, which exhibits a solid-angle
deficit, the Liouville solution is characterized by a solid-angle excess.
It corresponds to the spherical dislocation of the space-time within the
geometric theory of defects \cite{KatVol92,Katana05}.

We conclude that an observer at spatial infinity doesn't see the black hole, but
can confirm the presence of the geometric defect by measuring the area of a
large sphere. This is another essential geometric difference between the black
hole in the Liouville solution and the Schwarzschild one.

{\em Comparison with the Schwarzschild metric.}
Although the Liouville solution with the black hole differs considerably from
the Schwarzschild solution, a useful comparison can be made. Let us write
metric~(\ref{metrTR}) at large distances $R\to\infty$ and late time $T\to\infty$
assuming that $T/R\to 0$ to first order in $T/R$
\begin{equation} \label{aproxLiu}
  ds^2=dT^2-\frac{T}{R}dTdR-dR^2-4R^2d\Omega.
\end{equation}
This asymptotic is also valid for any finite $T$ if instead of condition
$T\to\infty$, one requires $c\to +0$.

The Schwarzschild solution in the Painlev\'e-Gullstrand coordinates
\cite{Peinle21,Gulldt22} has the form:
\begin{equation}\label{schwa}
  ds^2=dT^2-\left(dR+\sqrt{\frac{2M}{R}}dT\right)^2+R^2d\Omega,
\end{equation}
where $M=\text{const}>0$ is the mass of the Schwarzschild black hole. For
$R\to\infty$, metric~(\ref{schwa}) tends to the Lorentz metric, and, to the
linear order in $\sqrt{M/R}$, becomes:
\begin{equation*} \label{aproxSch}
  ds^2=dT^2-2\sqrt{\frac{2M}{R}}dTdR-dR^2-R^2d\Omega.
\end{equation*}
Comparing this expression with Eq.~(\ref{aproxLiu}), we get the Schwarzschild
mass function
\begin{equation}                                                   \label{masss}
  M(T,R)=\frac{T^2}{8R}.
\end{equation}
This expression was obtained under assumptions $R\to\infty$, $T\to\infty$, and
$T/R\to 0$ or $c\to0$. So, it corresponds to large distances and
sufficiently late time $T\gg c$. The event horizon equation~(\ref{evhor})
written without any approximation takes the standard Schwarzschild form
\begin{equation*}
  R=2M(T,R).
\end{equation*}

In terms of the mass function, metric (\ref{metrTR}) is
\begin{equation*}
  ds^2=\left(1+\frac c{2R}-\frac{2M}{R}\right)
  \left[\left(dT-\sqrt{\frac{2M}R}dR\right)^2-dR^2-4R^2d\Omega\right].
\end{equation*}
Note that the mass function can be finite in the considered limit.

The mass function (\ref{masss}) was obtained in the Painlev\'e-Gullstrand
coordinates, and naive transition to Schwarzschild coordinates
\begin{equation*}
  ds^2=\left(1-\frac{T^2}{4R^2}\right)dT^2-\frac{dR^2}{1-\frac{T^2}{4R^2}}-
  4R^2d\Om
\end{equation*}
looks nice but needs justification because transition from Schwarzschild to
Painlev\'e-Gullstrand coordinates requires $M=\text{const}$, and there is no
coordinate transformation that brings metric~(\ref{metrTR}) to this form.

{\em Conclusion.}
It is shown that the Liouville solution in General Relativity
with a scalar field and the exponential potential describe the formation of a
spherically symmetric black hole by a scalar field. The solution is global,
infinitely smooth, and invariant under global Lorentz transformations.
The space-time is conformally flat, and, hence, it's casual structure coincides
with the Minkowskian one.  This circumstance allows us to determine rather
straightforwardly the spacelike black hole singularity and the event horizon
which is the future light cone with the vertex at the origin of the coordinate
system. The horizon is a sphere which expands from the point to infinity with
the velocity of light. The spacelike singularity emerges after the event
horizon appears and is completely hidden by the horizon.

An interesting feature is that an external observer does not see the singularity
itself but can, in principle, observe the appearance of the horizon by measuring
the area of a large sphere surrounding it. The difference from the standard area
is due to the excess of the solid angle which shows the appearance of the
geometric defect at the origin.

At large distances, the Liouville solution is compared with the Schwarzschild
solution in the Painlev\'e-Gullstrand coordinates, and the mass function is
obtained.

The Liouville solution is simple and seems to be interesting. It raises too
many questions which, hopefully, will be answered in future.

D.~E. Afanasev and M.~O. Katanaev contributed equally to this work. The authors of this paper were ordered alphabetically.

\end{document}